\documentclass[final,5p,times,twocolumn]{elsarticle}

\usepackage{graphicx}  
\usepackage{amssymb}   
\usepackage{amsmath}
\usepackage{hyperref}
\usepackage{verbatim}
\usepackage{multirow}
\usepackage{float}
\usepackage{subfigure}
\usepackage[modulo]{lineno}
\usepackage{xcolor}

\journal{Nuclear Instruments and Methods in Physics Research Section B: Beam Interactions with Materials and Atoms}

\begin{document}
\begin{frontmatter}
\title{Mean range bunching of exotic nuclei produced by in-flight fragmentation and fission - Stopped-beam experiments with increased efficiency}



\cortext[cor1]{Corresponding author}
\author[1,2]{Timo Dickel\corref{cor1}}
\ead{t.dickel@gsi.de}
\author[1,2]{Christine Hornung}
\author[2]{Daler Amanbayev}
\author[1,2]{Samuel Ayet San Andr\'{e}s}
\author[1,2]{S\"onke Beck}
\author[2]{Julian Bergmann}
\author[1,2]{Hans Geissel}
\author[1]{Jürgen Gerl}
\author[1]{Magdalena G\'{o}rska}
\author[2]{Lizzy Gr\"of}
\author[1]{Emma Haettner}
\author[1,9]{Jan-Paul Hucka}
\author[1,2]{Daria A. Kostyleva}
\author[2]{Gabriella Kripko-Koncz}
\author[1,2,8]{Ali Mollaebrahimi}
\author[1]{Ivan Mukha}
\author[1]{Stephane Pietri}
\author[1,2]{Wolfgang R.\ Pla\ss}
\author[19]{Zsolt Podoly\'{a}k}
\author[1]{Sivaji Purushothaman}
\author[7,2]{Moritz Pascal Reiter}
\author[1,9]{Heidi Roesch}
\author[1,2,16]{Christoph~Scheidenberger}
\author[1,18]{Yoshiki K. Tanaka}
\author[1]{Helmut Weick}
\author[1]{Jianwei Zhao}
\author{and the Super-FRS Experiment Collaboration}

\affiliation[1]{
organization={GSI Helmholtzzentrum für Schwerionenforschung GmbH},
addressline={Planckstraße 1},
city={Darmstadt},
postcode={64291},
country={Germany}}

\affiliation[2]{
organization={II. Physikalisches Institut, Justus-Liebig-Universität},
addressline={Heinrich-Buff-Ring 16},
city={Gie{\ss}en}, 
postcode={35392},
country={Germany}}





\affiliation[7]{
organization={University of Edinburgh},
addressline={Edinburgh},
city={EH9},
postcode={3FD},
country={United Kingdom}}

\affiliation[8]{
organization={Nuclear Energy Group, ESRIG, University of Groningen},
addressline={Zernikelaan 25},
postcode={9747 AA},
city={Groningen},
country={The Netherlands}}

\affiliation[9]{
organization={Technische Universität Darmstadt},
addressline={Karolinenpl. 5},
city={Darmstadt},
postcode={D-64289},
country={Germany}}







\affiliation[16]{
organization={Helmholtz Research Academy Hesse for FAIR (HFHF), GSI Helmholtz Center for Heavy Ion Research, Campus Gießen},
addressline={Heinrich-Buff-Ring 16},
city={Gie{\ss}en},
postcode={35392},
country={Germany}}


\affiliation[19]{
organization={Department of Physics, University of Surrey},
addressline={Guildford},
city={GU2},
postcode={7XH},
country={United Kingdom}}

\affiliation[18]{
organization={High Energy Nuclear Physics Laboratory, RIKEN},
addressline={2-1 Hirosawa, Wako},
city={Saitama},
postcode={351-0198},
country={Japan}}
             
\begin{abstract}
The novel technique of mean range bunching has been developed and applied at the projectile fragment separator FRS at GSI in four experiments of the FAIR phase-0 experimental program. Using a variable degrader system at the final focal plane of the FRS, the ranges of the different nuclides can be aligned, allowing to efficiently implant a large number of different nuclides simultaneously in a gas-filled stopping cell or an implantation detector. Stopping and studying a cocktail beam overcomes the present limitations of stopped-beam experiments. The conceptual idea of mean range bunching is described and illustrated using simulations. In a single setting of the FRS, 37 different nuclides were stopped in the cryogenic stopping cell and were measured in a single setting broadband mass measurement with the multiple-reflection time-of-flight mass spectrometer of the FRS Ion Catcher.
\end{abstract}

\begin{keyword}

range bunching \sep exotic nuclei \sep mean range bunching \sep in-flight separator \sep thermalized beams \sep FRS ion catcher \sep MR-TOF-MS

\end{keyword}

\end{frontmatter}


\section{Introduction} \label{sc_intro}
Projectile fragmentation and fission are the main production methods for exotic nuclei at in-flight separators \cite{Blumenfeld2013, Geissel2014}; several of these are worldwide in operation, or under construction \cite{Ma2021}. They allow access to the shortest-lived isotopes of all elements for a wide range of experiments \cite{NAKAMURA2017}. To study the ground and isomeric state properties of nuclei, these are stopped in thin (less than 0.5 g/cm$^2$) active stoppers, such as gas-filled stopping cells \cite{Wada2013}, Double-sided Silicon Strip Detectors (DSSD), or plastic scintillators. The range straggling of heavy ions produced at energies of a few hundred MeV/u can be several g/cm$^2$; thus, only a fraction of all nuclei that are produced and identified can be measured simultaneously in stopped-beam experiments. This limitation can be overcome by a new technique, mean range bunching, which is applicable to ions produced at relativistic energies. The conceptual idea, simulations, and results for this technique are discussed and shown in the following for the first time.

\section{Concept and simulations} \label{sc_exp}
\begin{figure*}[th]
	\centering
	\includegraphics[width=0.9\textwidth]{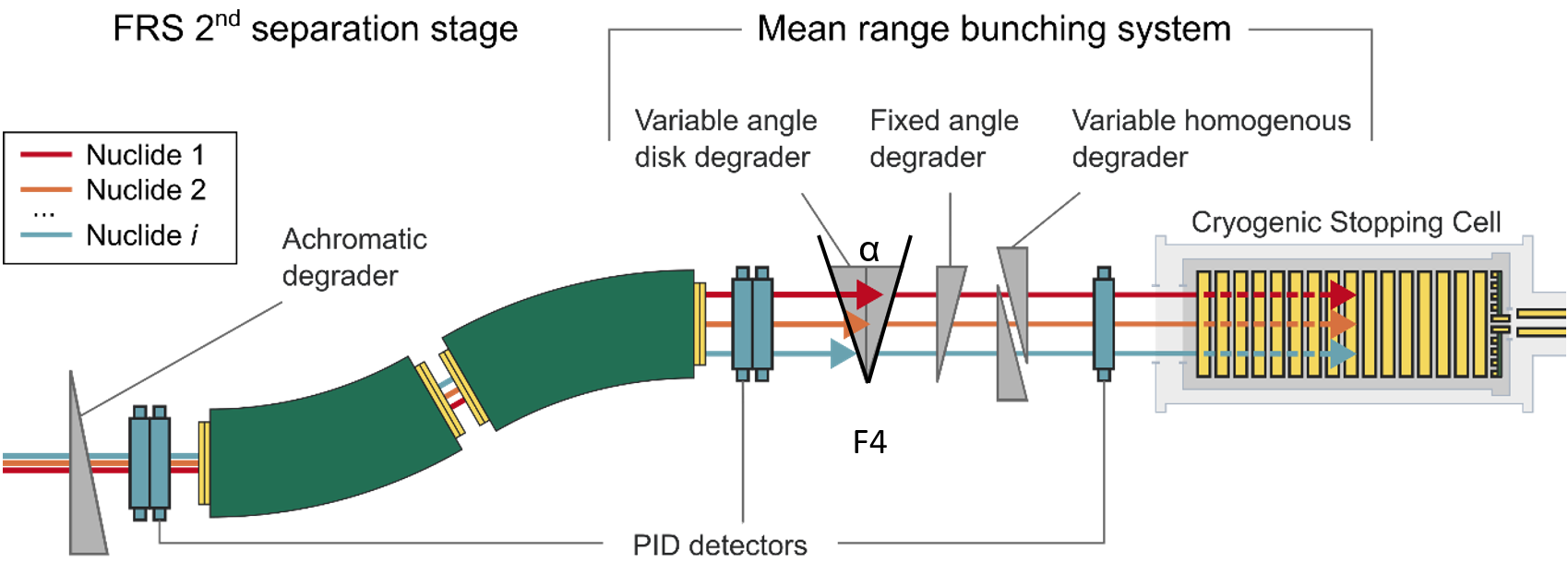}
	\caption{Schematic figure of the second half of the FRS, including the cryogenic stopping cell of the FRS Ion Catcher at the final focal plane (F4). Mean range bunching is conceptually shown for three different ion species. At the final focal plane, ions of different energies and without the wedge-shaped degraded with different ranges than the reference ions (orange) are indicated using different colors (red, blue). Passing through a wedge-shaped degrader system (variable angle disk degrader and fixed angle degrader) with the optimized angle $\alpha$ allows for aligning the range distributions of all nuclides and stopping all simultaneously in the cryogenic stopping cell. }
	\label{fig:mrb-scheme}
\end{figure*}
At the FRagment Separator FRS \cite{Geissel1992b} at GSI, exotic nuclei are produced, separated, and identified in-flight. In its standard ion-optical mode, the FRS uses the \mbox{$B \rho - \Delta E - B \rho$} method \citep{Geissel1992b} to achieve isotopic-spatial separation at the final focal plane. Experiments that require slowed-down ions in a gas-filled stopping cell or implantation detector benefit from the FRS operated in the mono-energetic mode \cite{Scheidenberger2003}. In this mode, the range straggling of ions of the same nuclide is minimized, maximizing the number of stopped ions of one species. However, ions of only one to three different nuclides can be stopped simultaneously since they are separated in range. Since the mean velocity of all nuclides in front of the degrader in the mid-focal plane is the same in the first order, it follows for the range $R$:
\begin{equation}
R \approx	\frac{A}{Z^2}.
\end{equation}
This separation in range exceeds the areal density of a stopping cell by two orders of magnitude. Thus, studying larger regions of the nuclide chart is time-consuming and requires a lot of precious beam time.

The Mean Range Bunching (MRB) technique has been developed to address this issue; see Fig.~\ref{fig:mrb-scheme}. Besides the development of the novel concept itself, the recently commissioned full degrader system at the final focus of the FRS was crucial to implement mean range bunching. For individual isotopes, the stopping efficiency with a very thin stopper, like a gas-filled stopping cell, is lower than in the mono-energetic mode. With mean range bunching, ions of many nuclides can be stopped simultaneously, rather than one at a time, in the cryogenic stopping cell (CSC) at the FRS Ion Catcher (FRS-IC) \cite{Plass2013}. In the same way, the method is applicable for broadband measurements of stopped cocktail beams in a thin DSSD active stopper for half-life and $\beta$-delayed proton branching-ratio measurements.

\begin{figure*}[tb] 
        \centering
        
   \subfigure{\includegraphics[width=0.47\linewidth]{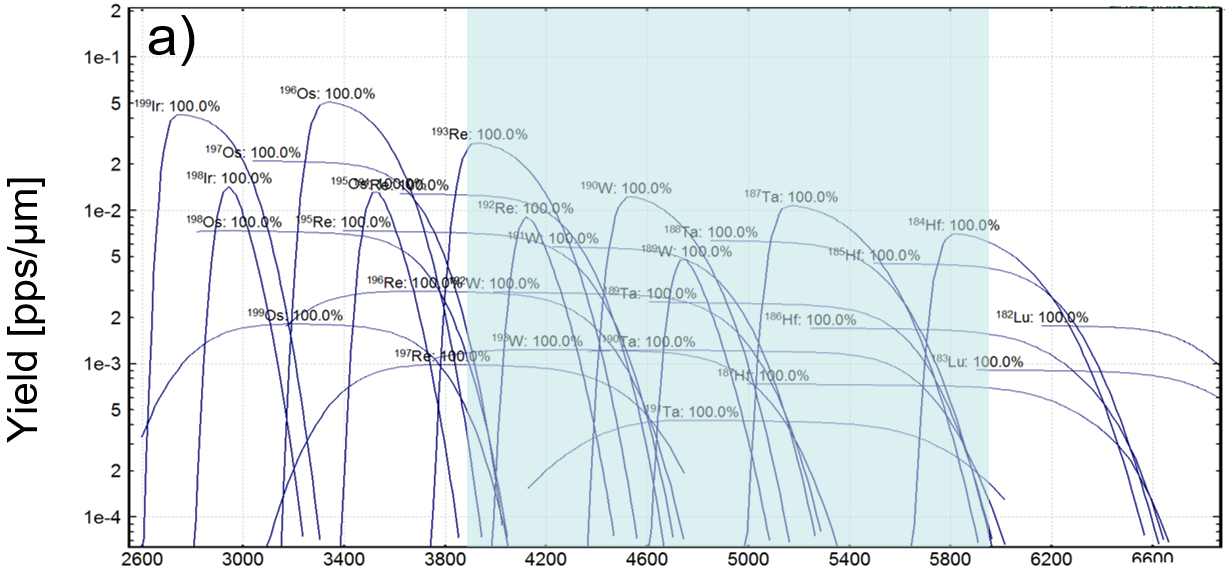}}\qquad
   \subfigure{\includegraphics[width=0.47\linewidth]{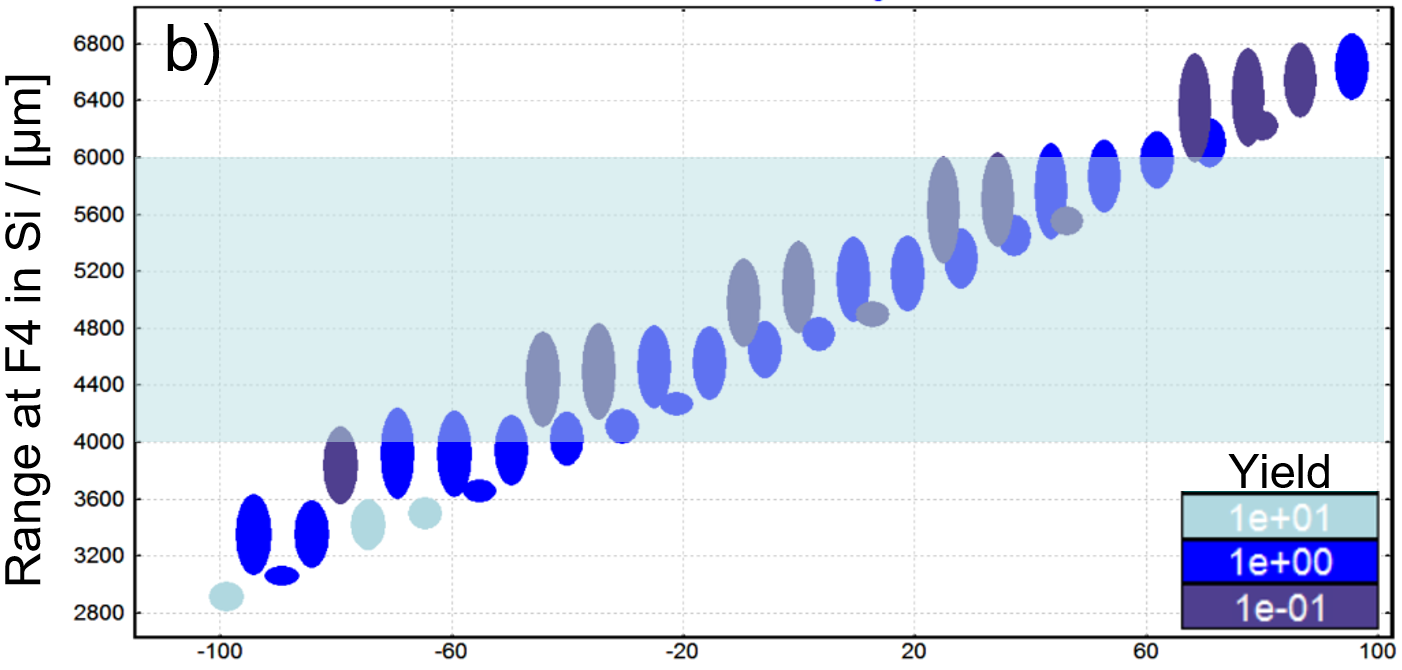}}\\
   \subfigure{\includegraphics[width=0.47\linewidth]{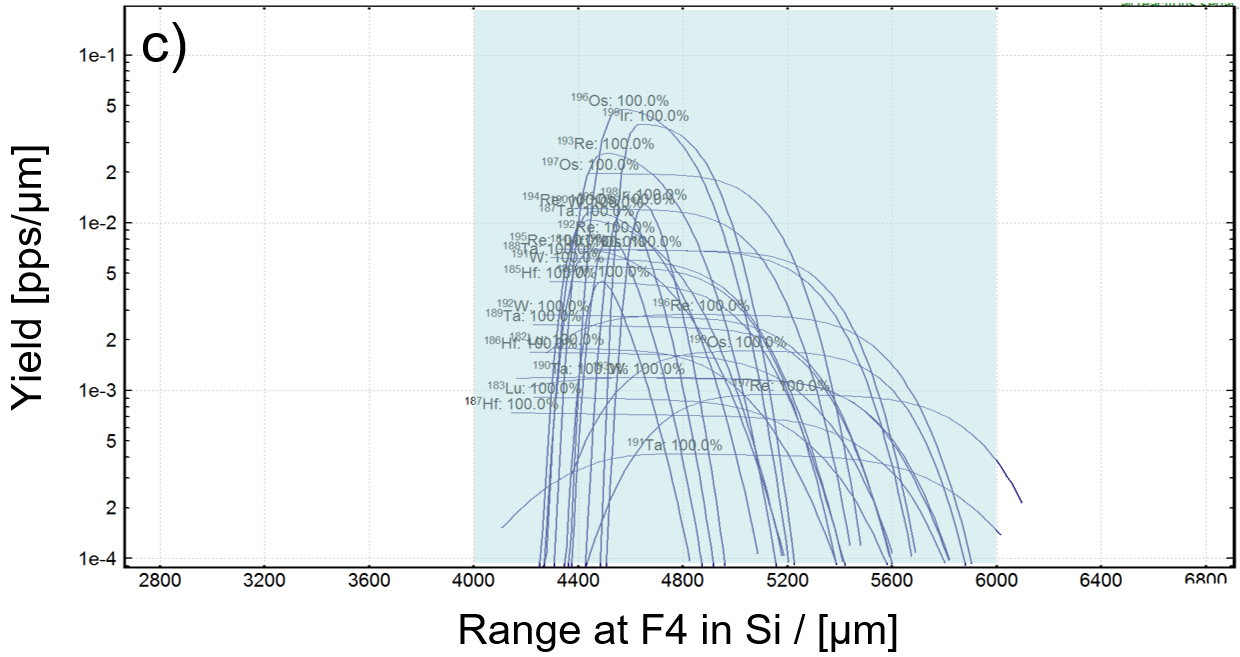}}\qquad
   \subfigure{\includegraphics[width=0.47\linewidth]{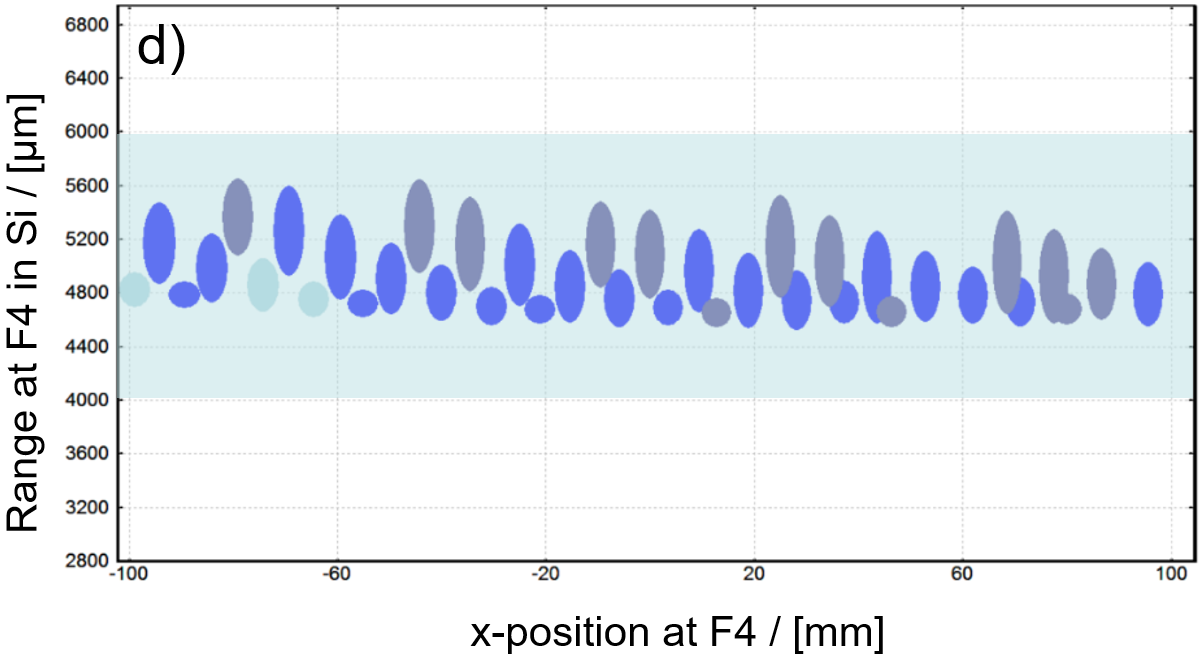}}

        \caption[Mass range bunching S450]{Simulation results of the residual range of ions for a decay spectroscopy campaign by the DESPEC collaboration around the neutron shell closure $N$=126 (S450), performed using LISE++ \cite{Tarasov2008}. A $^{208}$Pb beam with 1 GeV/u impinges on a 2500 mg/cm$^2$ Be Target (a, c) Stopping position in Si, representing the DSSDs of the AIDA detector \cite{MISTRY2022}. (b, d) 2D plot of range vs. position in the dispersive plane at the final focus of the FRS. The colour represents the yield.  (a, b) Standard operation: The range distribution is about 4 mm in Si. (c, d) Mean range bunching: the center of the stopping positions for different nuclides is compressed by almost an order of magnitude to about 0.6 mm in Si.}
        \label{fig:MRB_S452}
    \end{figure*}

When the FRS is used in achromatic mode, in contrast to the earlier explained mono-energetic mode, the range straggling of ions of the same nuclide is not minimized at the final focal plane. But in the achromatic ion optic, the ions of a particular nuclide are focused on a narrow spot at the final focus and thus penetrate through approximately the same amount of matter even when a wedge-shaped degraded is used in the focal plane. In this mode, the range of ions of different nuclides at the final focal plane of the FRS linearly depends on their lateral ($x$) position (dispersive plane) (Fig.~\ref{fig:MRB_S452} c)):
\begin{equation}
	\left(\frac{dR}{dx}\right)_i = \text{const},
\end{equation}
where $i$ indicates that the dependence is material-specific.
Passing the beams through an additional wedge-shaped degrader system (variable angle disk degrader and fixed angle degrader) at the final focus allows for aligning the range distributions of all nuclides. The degrader is oriented, so nuclei with a larger range penetrate through the thicker layers of matter. If the wedge angle~$\alpha$ of the degrader fulfills the condition
\begin{equation}
	\tan(\alpha) = \frac{1}{\rho_i} \cdot \left(\frac{dR}{dx}\right)_i,
\end{equation}
where $\rho_i$ is a volume density of the degrader material, the mean of the ranges of all nuclides become aligned and equal to the one of the nuclides at the central optical axis ($x$=0) (Fig.~\ref{fig:MRB_S452} d)). The variable homogeneous degrader adjusts the mean range and, thus, the stopping position to the goal volume. Due to the same mean range of all nuclides, all are now stopped simultaneously in a thin stopping cell or detector setup.  

Mean range bunching can be used for all experiments that benefit from stopped cocktail beams at GSI and FAIR as it reduces the required beam time for large-scale mapping of nuclear properties and searching for new isotopes and isomeric states. So far, mean range bunching has been used in four experiments with the FRS since 2020. Simulations for mean range bunching for a decay spectroscopy campaign by the DESPEC collaboration around the neutron shell closure $N$=126 are shown in Fig. ~\ref{fig:MRB_S452}. While with the standard achromatic mode of the FRS, the total width of the range distribution of all nuclides amounts to 4 mm in Si, and only about half of the nuclides can be stopped in the DSSDs of the AIDA detector, with mean range bunching the width of the total range distribution can be reduced by almost an order of magnitude, allowing to implant all nuclides in the detector. The method was successfully applied in four experiments: (i) two isotope search experiments connected with mass and half-life measurements at $N$=126 (S468) and in the region of neutron-deficient lanthanides (S482), (ii) an experiment to study fission isomers (S530) detected with the FRS Ion Catcher and a scintillator implanter and (iii) decay spectroscopy experiment around the neutron shell closure $N$=126.

\section{Experimental results} 
    In this section, exemplary results are shown, illustrating the potential and proof of the performance of the mean range bunching technique. The example given here is broadband mass measurements in neutron-deficient lanthanides. A degrader system that enables mean range bunching was installed at the final focal plane of the FRS. It has similarities to the system at the mid-focal plane of the FRS ~\cite{Geissel1992b}. Two wedge-shaped aluminum disks are rotated in opposite directions, and thus the wedge angle $\alpha$ along the $x$-direction can be adjusted. Two wedges form a homogeneous variable degraded. Two ladders are equipped with aluminum and tantalum homogeneous plates, and a fixed-angle aluminum wedge. The latter allows extending the angle $\alpha$ beyond the limits of the rotating disks. Similar fixed-angle wedges have recently also been installed at the mid-focal plane to enable the monochromatic operation of the FRS for light and neutron-rich heavy beams, especially for light fission fragments. The tantalum plates are used to reduce nuclear reaction losses during the slowdown, compared to lighter-Z stopping materials. The stopped beam rate can be increased by up to 30\% when replacing a significant fraction of the usual aluminum degraded with tantalum \cite{Groef2023}. 
    
    At the final focal plane, behind the new degrader system, the FRS-IC is installed (Fig.~\ref{fig:mrb-scheme}). The FRS-IC consists of three major components: (i) the CSC \cite{Ranjan2015, Purushothaman2013} for efficiently slowing down and thermalization of the short-lived exotic nuclei produced at relativistic energies, (ii) a Radio-Frequency Quadrupole beamline \cite{Reiter2015, Haettner2018} for mass-selective transport, and differential pumping, (iii) a multiple-reflection time-of-flight mass spectrometer (MR-TOF-MS) \cite{Dickel2015b,Plass2008} for performing direct mass measurements. The MR-TOF-MS combines a unique set of performance parameters - fast ($\sim$ms) \cite{Ayet2019}, accurate \cite{Mardor2021}, broadband and non-scanning \cite{Dickel2015b} operation, and isobar and isomer separation \cite{Dickel2015}. The mass measurements are done by injecting ions into the isochronous Time-Of-Flight (TOF) analyzer of the MR-TOF-MS, where they undergo many turns between two electrostatic mirrors. The ions are ejected from the TOF analyzer by opening the exit mirror toward the detector to record the time of flight for the ions.    
    
    Mean range bunching was shown for mass measurements of neutron-deficient lanthanides produced via projectile fragmentation of a 1050~MeV/u $^{208}$Pb projectile beam. A primary beam with an intensity of up to $6\cdot 10^8$ ions per spill, with a typical spill length of $\mathrm{1.5~s}$ and a repetition rate of about 0.4~Hz on a beryllium production target at the entrance of the FRS with an areal density of $\mathrm{3.991~g/cm^2}$ and a $\mathrm{0.225~g/cm^2}$ niobium stripper, was used. The FRS was set up with achromatic optics with a $\mathrm{2 g/cm^2}$ aluminum degraded with a wedge handle of 2.45 mrad at the mid-focal plane. The centered fragment was fully stripped $^{139}$Dy. The slits behind the target were closed to $\pm$8.6~mm, and the slits at the first focal plane were -10/+30~mm. The slits at the final focal plane were opened completely. In order to apply mean range bunching, i.e.\ to align all nuclides in range at the final focal plane, a variable homogeneous degrader with a thickness of $\mathrm{1682.1~mg/cm^2}$, a fixed-angle degrader with a central thickness of $\mathrm{1080.8~mg/cm^2}$ with a slope of $\mathrm{17.5~mrad}$ and a variable-angle disk degrader of $\mathrm{992.2~mg/cm^2}$ and an angle of $\mathrm{5.5~mrad}$ were used.
    
\begin{figure} 
        \centering
        {\hspace{-0.0 cm}\includegraphics[width=1\linewidth]{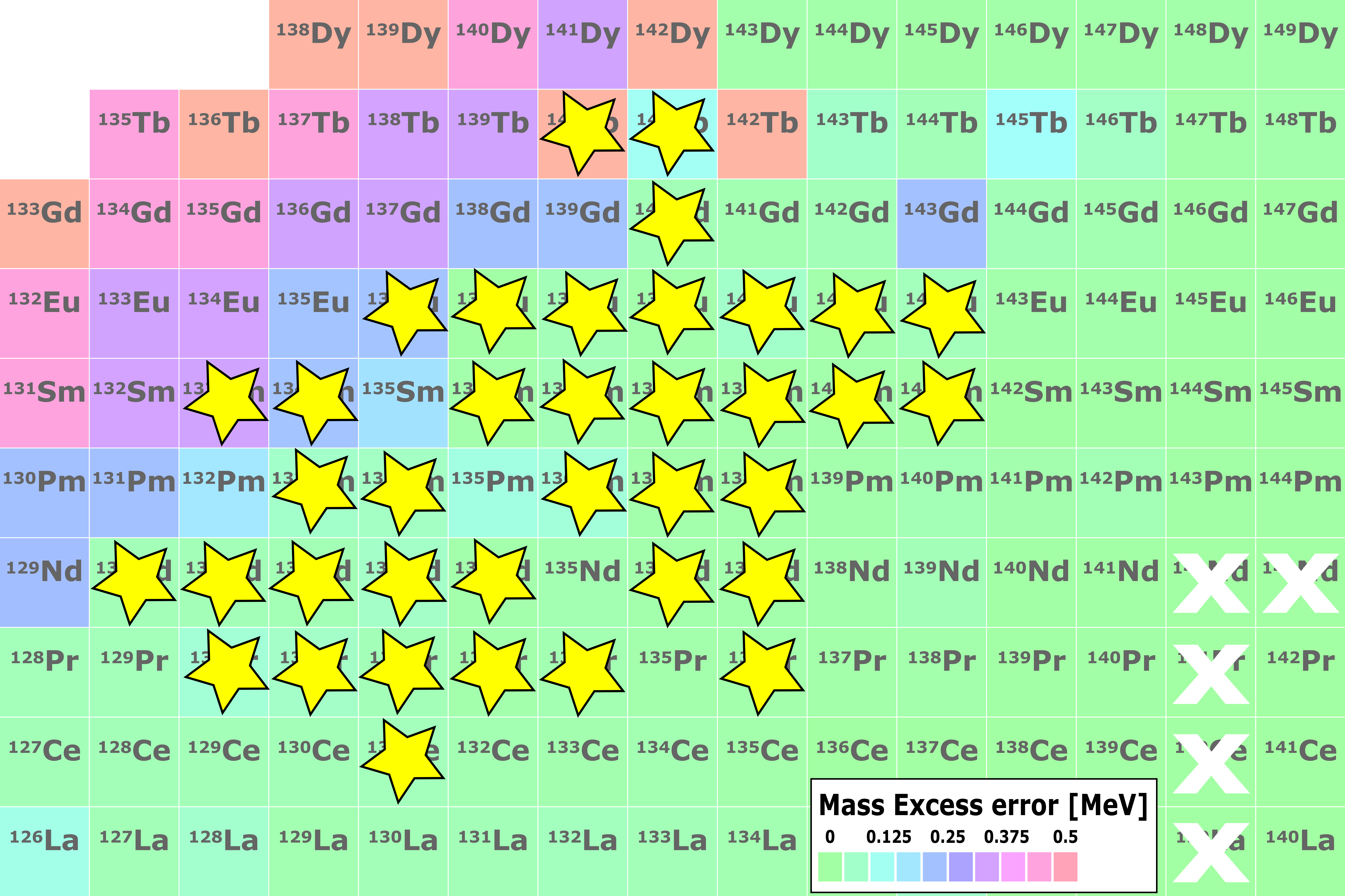}}
        \caption[S482]{Mapping the mass surface for the neutron-deficient lanthanides with the FRS Ion Catcher while employing mean range bunching. The figure shows a section of the chart of the nuclides in the range from La to Dy. Stable nulcides are marked with a white X. The 37 nuclides, which were measured in a single setting of the FRS and in a single setting of the MR-TOF-MS are marked by stars.}
        \label{fig:S482}
    \end{figure}

    The ions stopped in the CSC were transported to the MR-TOF-MS and identified by their mass-to-charge ratios. In the TOF-analyzer of the MR-TOF-MS, the ions underwent, depending on their mass, 300 to 320 isochronous turns; despite this broadband setting of the device, a mass resolving power of 350,000 was achieved. In the measurement, more than 37 nuclides were stopped and measured simultaneously; see Fig.~\ref{fig:S482}. Eight elements were covered from Ce to Tb, and a mass range from 130~u to 142~u. The ions of mass 135~u were also stopped but could not be measured in this setting of the MR-TOF-MS since they were too close to the exit mirror when it was opened, exposing them to switching electric fields. The mean range bunching technique ideally complements the broadband capabilities of the MR-TOF-MS of the FRS Ion Catcher. In a similar way it can enhance the measurement capabilities of other experiments with thin stoppers.

    It should be noted that an essential advantage of the mean range bunching technique is that it employs the FRS in achromatic mode, the most often used operation mode. Thus it allows stopped beam experiments (e.g., mass, half-live and decay spectroscopy measurements), in a symbiotic mode with experiments such as searches for new isotopes simultaneously. Instead of bunching, the method can also be used to increase the mean range distribution, i.e., mean range de-bunching. This can be used to run two experiments in parallel with different stopped isotopes. Experiments that are envisaged here are, for example, the combination of the Super-FRS Ion Catcher \cite{Dickel2016} and the DESPEC setup \cite{MISTRY2022}.

\section{Conclusions}
    A method, mean range bunching, has been developed to stop a cocktail beam of relativistic, in-flight separated exotic nuclei efficiently in a thin stopper. It was applied in several experiments at the projectile fragment separator FRS at GSI in FAIR Phase-0. The method takes advantage of the correlation of horizontal position and  the residual range of the in-flight separated ions at the final focal plane. With a recently installed variable degrader system at the final focal plane of the FRS, the ranges of the different nuclides can be aligned, allowing to efficiently implant a large number of different nuclides simultaneously in a gas-filled stopping cell or a thin decay detector (thin means: areal weight of the stopper is small as compared to the range distribution and in particular to the range differences of the ions of interest), thus boosting the reach of the experiment and allowing for efficient broadband measurements. Mean range bunching has been demonstrated for mass measurements of light neutron-deficient lanthanides with the FRS Ion Catcher. In a single setting of the FRS, 37 different nuclides could be stopped in the CSC and measured in a single setting of the MR-TOF-MS of the FRS-IC. The development of mean range bunching will have large implications for all stopped beam experiments at the FRS at GSI and in the future at the Super-FRS at FAIR, as it tremendously reduces the required beam time for large-scale mapping of nuclear properties. Its use is not limited to GSI and FAIR, but is applicable at other in-flight separators, too.

\section*{Acknowledgments}
    The results presented here are based on the experiments S450, S468, S482, and S530, which were performed at the FRS at the GSI Helmholtzzentrum f{\"u}r Schwerionenforschung, Darmstadt (Germany) in the context of FAIR Phase-0.
    This work was supported by the German Federal Ministry for Education and Research (BMBF) under contracts no.\ 05P19RGFN1 and 05P21RGFN1, by the German Research Foundation (DFG) under contract no.\ SCHE 1969/2-1, by the Helmholtz Research Academy Hesse for FAIR (HFHF), by HGS-HIRe, and by Justus-Liebig-Universit{\"a}t Gie{\ss}en and GSI under the JLU-GSI strategic Helmholtz partnership agreement.





\bibliographystyle{apsrev4-1}
\bibliography{Ref-EMIS.bib}







\end{document}